\documentclass[10pt,letterpaper,twocolumn]{article} %% two column, final layout

\usepackage{ol2}
\usepackage[draft]{hyperref}
\usepackage{amsmath}

\begin{document}

\twocolumn[ %% activate for two-column option

\title{Chiral excitation and effective bandwidth enhancement in tilted coupled optical waveguide lattices}

%% For REVTeX it is possible to automate superscript and e-mail callouts with the superscriptaddress option; see REVTeX4 documentation.

\author{Stefano Longhi}
\address{Dipartimento di Fisica, Politecnico di Milano and Istituto di Fotonica e Nanotecnologie del Consiglio Nazionale delle Ricerche, Piazza L. da Vinci 32, I-20133 Milano, Italy (stefano.longhi@polimi.it)}
\address{IFISC (UIB-CSIC), Instituto de Fisica Interdisciplinar y Sistemas Complejos, E-07122 Palma de Mallorca, Spain}

\begin{abstract}
Light escape from an optical waveguide side-coupled to a waveguide lattice provides a photonic analogue of
the spontaneous emission process of an excited two-level atom in a one-dimensional array of cavities.
 According to the Fermi golden rule the decay process is prevented when the atomic resonance frequency
falls in a stop band of the lattice, while time-reversal symmetry ensures that the spontaneously emitted photon has equal probability to propagate in opposite directions of the array. This scenario is drastically modified when the quantum emitter drifts along the lattice at a constant speed. In the waveguide optics analogue the atomic drift is emulated by the introduction of a slight geometric tilt of the waveguide axis from the lattice axis.
 In this setting light excitation in the array is chiral, i.e. light propagates in a preferred direction of the lattice, and coupling is allowed even though the waveguide is far detuned from the tight-binding lattice band.
\end{abstract}

%\ocis{130.2790,  130.3120, 000.160)}
 ] %% activate for two-column option

{\it Introduction.}  
Propagation of discretized light by evanescent mode coupling plays
a central role in integrated classical and quantum photonics \cite{r1,r2,r3,r4,r4b}, providing a useful platform for simulating a plethora of coherent quantum phenomena in the matter \cite{r5}. 
A prototypal system is provided by one (or more) waveguide side-coupled to a 
waveguide lattice, where light escape dynamics into the lattice emulates the quantum mechanical decay process of one (or more) quantum emitter into a tight-binding continuum \cite{r5,r5b,r6,r7,r8,r9}.
 The optical analogues of important phenomena like non-exponential quantum decay, Zeno and anti-Zeno dynamics, Fano interference, dark states, giant atom decay, superradiance dynamics, and decay in a hyper-continuum  have been reported in such photonic platform \cite{r5,r6,r7,r8,r9,r10,r11,r12,r13,r14,r15,r16,r17}.  In most of such previously investigated systems, the decay dynamics is described by an Hamiltonian possessing time-reversal symmetry, which prevents  chiral emission in the lattice, i.e. light is equally coupled in opposite directions of the lattice. In fact, chiral emission requires rather generally the use of synthetic gauge fields to break time reversal symmetry or additional degrees of freedom that ensure directional coupling \cite{r18,r19,r20,r21,r22,r23,r24,r24a}.
 % Also phase matching condition (corresponding to energy conservation in the quantum mechanical problem) permits light coupling into the lattice provided that the propagation constant mismatch of the side waveguide from the guides in the array is smaller that the band of the tight-binding lattice. 
 Also, in the weak coupling limit the rate of light escape is ruled by the Fermi golden rule \cite{r7,r8}, which predicts a vanishing decay rate when the density of states in the continuum vanishes (lattice stop band). \\
In this Letter it is shown that such a common scenario of evanescent field coupling drastically changes when a waveguide is side-coupled to a {\it tilted} waveguide lattice, which would correspond in the quantum mechanical problem to a quantum emitter moving at a constant speed along the lattice. In such an optical setting, light excitation in the array is chiral and the waveguide-lattice coupling occurs even though the propagation constant mismatch is far beyond the tight-binding lattice band, resulting in an effective enhancement of the lattice band.\\ 
%Chirality and bandwidth enlargement are explained by an effective tilting of the tight-binding lattice dispersion curve.\\ 
 \par
 {\it Photonic system and quantum-optical analogy.} Let us consider the photonic system schematically depicted in Fig.1(a), consisting of a straight dielectric optical waveguide W which is side-coupled  to an infinitely-extended one-dimensional waveguide lattice. The waveguide W is vertically displaced by a distance $h$ from the array and its optical axis is tilted, with respect to the optical axis $z$ of the lattice, by a small angle $\alpha$ , so that the evanescent field coupling of W with the various guides in the lattice changes with $z$. Photon propagation in the system is described by the tight-binding Hamiltonian (see e.g. \cite{r4,r17,r24b})
 \begin{equation}
 \hat{H}=\hat{H}_a+\hat{H}_{b}+\hat{H}_{int}
 \end{equation}
 \begin{figure}[htb]
\centerline{\includegraphics[width=8.7cm]{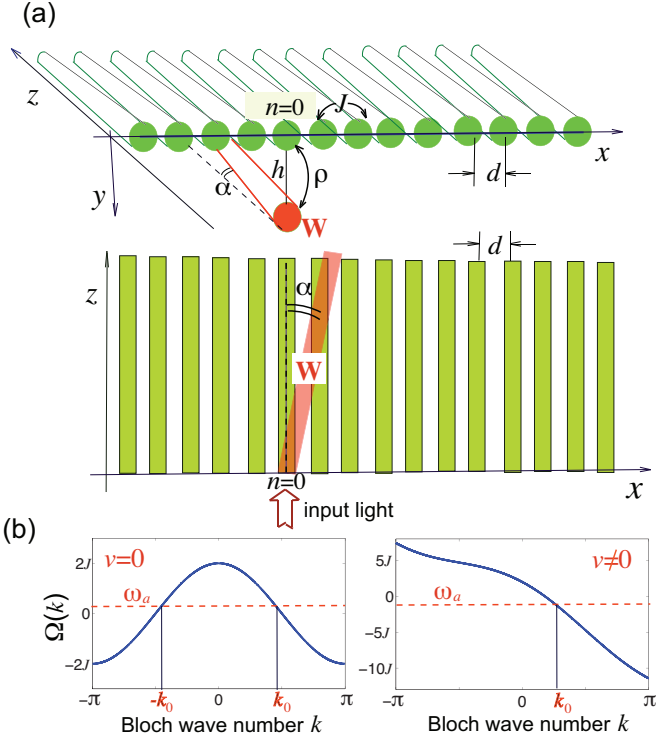}} \caption{ \small
(Color online) (a) Schematic of the integrated photonic system. An optical waveguide W is side-coupled to a one-dimensional waveguide lattice and tilted by an angle $\alpha$ along the propagation direction $z$. $d$ is the lattice period, $h$ the vertical displacement of waveguide W from the array, $J$ and $\rho$ are the coupling constants. Light escape dynamics from waveguide W emulates the spontaneous emission process of a moving two-level atom into a tight-binding bath. (b) Effective dispersion curve $\Omega(k)=2 J \cos k-vk$ of the lattice for $v=0$ (left panel), and $v=3J$ (right panel).}
\end{figure} 
 where $\hat{H}_a= \omega_a \hat{a}^{\dag} \hat{a}$ is the Hamiltonian of the photon field in waveguide W,  $\hat{H}_b=  J \sum_n ( \hat{b}^{\dag}_n \hat{b}_{n+1}+ H.c.)$ is the tight-binding Hamiltonian of the photonic modes  of the lattice, and 
 \begin{equation}
 \hat{H}_{int}= \sum_{n}  \rho(n-vz) ( \hat{a}^{\dag} \hat{b}_{n}+H.c.)
 \end{equation}
  is the Hamiltonian describing evanescent mode coupling of waveguide W with the lattice. In the above equations, $\hat{a}^{\dag}$ and $\hat{b}^{\dag}_n$ are the bosonic creation operators of photons in waveguide W and in the $n$-th waveguide of the lattice, respectively, satisfying the usual commutation relations $ [\hat{a}, \hat{a}^{\dag}]=1$, $[ \hat{b}_l, \hat{b}^{\dag}_n]= \delta_{n,l}$, etc.; $J$ is the coupling constant between adjacent waveguides in the lattice; $\omega_a$ is the propagation constant shift of waveguide W from the propagation constant of the lattice waveguides (also accounting for the tilt-induced shift); $\rho(n-vz)$ is the coupling constant between W and the $n$-th lattice waveguide; $v=\alpha /d $ is the drift parameter introduced by the tilt angle $\alpha$; and $d$ is the lattice period [Fig.1(a)]. Since the coupling constant $\rho$ between two waveguides spaced by a distance $r$ is well described by an exponential decreasing function of $r$, namely $\rho=\rho_0 \exp(-\gamma r)$ with $\rho_0$ and $\gamma$ constant parameters \cite{r2,r25}, the $z$-varying coupling factor $\rho(n-vz)$ entering in Eq.(2) reads 
   \begin{equation}
 \rho(n-vz)=\rho_0 \exp\left[- \gamma \sqrt{h^2+d^2(n-vz)^2} \right].
 \end{equation} 
 In a typical geometrical setting, $\rho(x)$ is a bell-shaped function which takes the largest value at $x=0$ and rapidly decreases as $|x|$ increases, with a full-width at half maximum $\Delta x$ of the order $\Delta x  \sim 1-3$ [Fig.2(a)]. 
 As discussed in previous works \cite{r16,r17}, in the single excitation sector of Fock space the escape dynamics of a photon from waveguide W into the array effectively emulates the spontaneous emission process of a two-level atom coupled to a one-dimensional array of optical cavities, where $z$ is time and $\omega_a$ is the frequency detuning between the atomic resonance frequency and the frequency of the cavity modes. For a non-vanishing value of $v$, the atom-array coupling periodically changes in time, as if the two-level atom slowly moves along the lattice at the speed $v$. Thus our photonic platform can emulate the radiative emission of a quantum emitter moving on a lattice, which shows interesting features as discussed below.\\
 \par
 {\it Light escape dynamics.} To describe photon escape dynamics in waveguide W, it is worth switching from Wannier to Bloch basis representation \cite{r17} and to consider a reference frame of the bosonic destruction operators rotating at the offest $\omega_a$. After introduction of the bosonic operators 
 \begin{eqnarray}
 \hat{A} & \equiv & \hat{a} \exp(i \omega_a z) \\
 \hat{C}(k) & \equiv & \frac{1}{ \sqrt{2 \pi}}  \left( \sum_n \hat{b}_n \exp(-ikn)  \right) \exp(i \omega_a z+ikvz)
 \end{eqnarray}
 where $-\pi \leq k < \pi$ is the Bloch wave number, the full Hamiltonian of the photon field is given by Eq.(1) with
\begin{eqnarray}
\hat{H}_a & = & \omega_a \hat{A}^{\dag} \hat{A} \nonumber \\
\hat{H}_b & = & \int_{-\pi}^{\pi} dk \; \omega(k) \hat{C}^{\dag}(k) \hat{C}(k) \\
\hat{H}_{int} & = & \int_{-\pi}^{\pi} dk \left\{ G(k,z) \hat{A}^{\dag} \hat{C}(k) +H.c.  \right\} \nonumber
\end{eqnarray}
where 
\begin{equation}
\omega(k) \equiv 2 J \cos k
\end{equation}
is the dispersion relation of the tight-binding energy band and $G(k,z)$ is the $z$-dependent spectral coupling function.  
$G(z,k)$ turns out to be periodic in $z$ with period $1/v$, and its Fourier decomposition reads
\begin{equation}
G(k,z)=\sum_l G_l(k) \exp( 2 \pi i l v z)
\end{equation}
with harmonic coefficients given by
\begin{equation}
G_l(k)=\frac{1}{\sqrt{2 \pi}}\int_{-\infty}^{\infty} dx \rho(x) \exp[ i(k+2 l \pi) x]
\end{equation} 
 ($l=0, \pm1, \pm2,...$, $-\pi \leq k < \pi$). For typical parameters values, only the zero-order term (mean value) in the Fourier series is non-negligible, i.e, at leading order one can assume 
 \begin{equation}
 G(k,z) \simeq G_0(k)=\frac{1}{\sqrt{2 \pi}}\int_{-\infty}^{\infty} dx \rho(x) \exp (ik x)
 \end{equation}
  independent of $z$ [see Fig.2(b)]. Note that, under such an assumption the spectral coupling function is the Fourier integral of the coupling factor $\rho(x)$, given by Eq.(3).\\
  The Heisenberg equations of the operators $\hat{A}$ and $\hat{C}(k)$ for the photon fields in waveguide W and in the array read
  \begin{eqnarray}
  i \frac{d \hat{A}}{dz} & = & - \omega_a \hat{A}+ [ \hat{A}, \hat{H}] \\
  i \frac{d \hat{C}}{dz} & = & - (\omega_a+kv) \hat{C}+ [ \hat{C}, \hat{H}] .
  \end{eqnarray} 
  After substitution of Eqs.(1) and (6) into Eqs.(11) and (12), one obtains
  \begin{eqnarray}
  i \frac{d \hat{A}}{dz} & = & \int_{-\pi}^{\pi} G_0(k) \hat{C}(k) \\
  i \frac{d \hat{C}}{dz} & = & (\Omega(k)-\omega_a ) \hat{C}+ G^*_0(k)  \hat{A}
  \end{eqnarray} 
  where we have set
  \begin{equation}
  \Omega(k) \equiv \omega(k)-kv=2J \cos k-kv.
  \end{equation} 
  Note that Eqs.(13) and (14) reduce to the usual $c$-number coupled-mode equations of light transport in the waveguide-lattice system when dealing with a single photon or a classical (coherent) light field \cite{r24b}.  
 \begin{figure}[htb]
\centerline{\includegraphics[width=8.7cm]{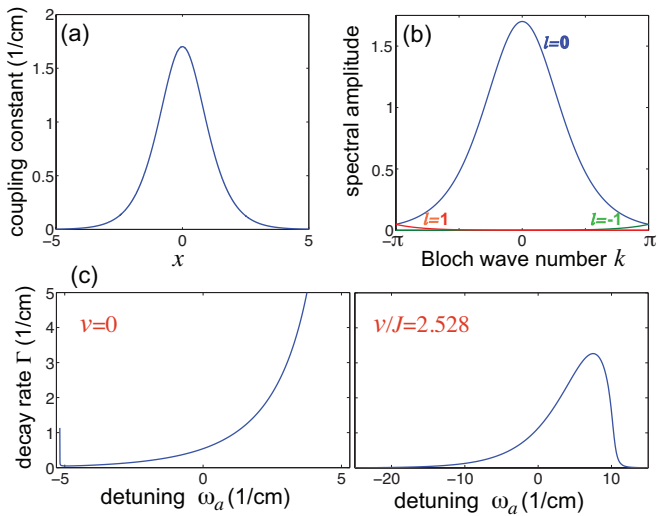}} \caption{ \small
(Color online) (a) Typical behavior of the coupling constant factor $\rho(x)$, given by Eq.(3), for parameter values $\kappa_0= 13.89 \; {\rm cm}^{-1}$, $\gamma=0.14 \; \mu {\rm m}^{-1}$, $h=15 \; \mu {\rm m}$ and  $d=12 \; \mu {\rm m}$ (corresponding to  a lattice coupling constant $J \simeq 2.59 \; {\rm cm}^{-1}$). Such physical parameters correspond to waveguide arrays manufactured by the femtosecond laser writing technique in fused silica and probed at 633 nm photon wavelength \cite{r25}. (b) Numerically-computed spectral amplitudes $|G_l(k)|$ for lower-order Fourier terms ($l=0, \pm1$). (c) Decay rate $\Gamma$ [Eq.(17)] versus detuning $\omega_a$ for two values of the angle $\alpha$.}
\end{figure} 
\begin{figure}[htb]
\centerline{\includegraphics[width=8.7cm]{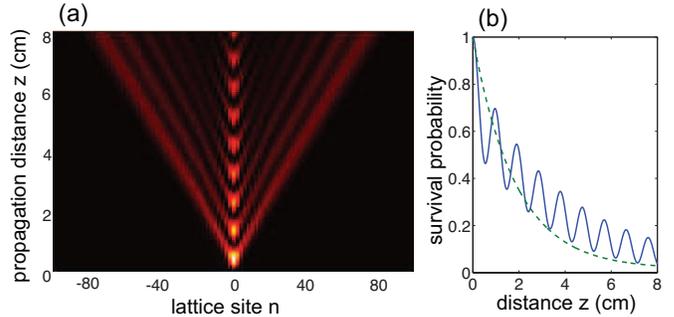}} \caption{ \small
(Color online) Light escape dynamics in the waveguide-array system of Fig.1(a) for parameter values as in Fig.2 and for $\omega_a=0$, $\alpha=0$. (a) Snapshot on a pseudo color map of the light intensity distribution in the lattice versus propagation distance. (b) Behavior of the photon survival probability $P_s(z)$ in waveguide W (solid curve), and exponential decay law predicted in the weak coupling limit (dashed curve). Light excitation in the array is bidirectional, as clearly manifested by the symmetric discrete diffraction pattern at around $n=0$. The oscillations in the decay of $P_s(z)$ are mainly due to the strong and non-local coupling of waveguide W with the array, resulting in damped Rabi-like oscillations.}
\end{figure} 
\begin{figure}[htb]
\centerline{\includegraphics[width=8.7cm]{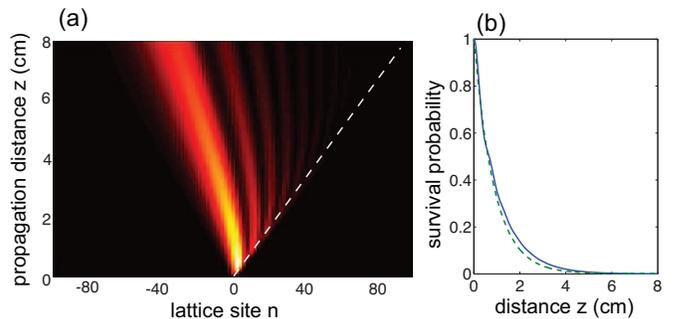}} \caption{ \small
(Color online) Same as Fig.3, but for $\alpha= \pi /400$, corresponding to $v /v_c \simeq 1.264$. In (a) the dashed line shows the axis of the tilted waveguide W. Note that light excitation in the array is chiral, with waves propagating backward in the lattice (crossing point $k_0>0$, corresponding to a negative group velocity $v_g= -2 J \sin k_0$). }
\end{figure} 
Let us now assume that at the initial plane $z=0$ a photon is injected into waveguide W. To physically understand the key role of a non-vanishing drift term $v$ on the photon escape dynamics, 
let us consider the weak coupling limit ($G \rightarrow 0$), i.e $J \gg \rho$,  and neglect possible retardation effects arising from non-local coupling \cite{r17}. Then a non-negligible probability per unit time that the photon tunnels into the lattice occurs at the Bloch wave numbers satisfying the resonance condition
\begin{equation}
\Omega(k)= \omega_a
\end{equation}
and the survival probability $P_s(z)$, i.e.the probability that the photon remains in waveguide W, decays following a nearly exponential law, $P_s(z) \simeq \exp(-\Gamma z)$, with a decay rate $\Gamma$ given by the Fermi golden rule
\begin{eqnarray}
\Gamma & = & 2 \pi \int_{-\pi}^{\pi} dk |G_0(k)|^2 \delta (\Omega(k)-\omega_a) \nonumber \\
& = &  2 \pi \sum_{k_0} \frac{|G_0(k_0)|^2}{ \left| v+2 J \sin k_0  \right| } 
\end{eqnarray}
where the sum on the right hand side of Eq.(17) is extended over all the roots $k_0$ of Eq.(16). Typical behavior of the decay rate versus $\omega_a$ is shown in Fig.2(c).
The propagation direction of the photon in the array is determined by the sign of the group velocity $v_g=(d \omega / dk)=-2 J \sin k$ at $k=k_0$. Let us distinguish two cases:\\
(i) First, let us assume $\alpha=0$, i.e. a zero drift velocity $v=0$. In this case, from Eq.(16) it follows that photon decay arises provided that $|\omega_a|< 2 J$, i.e. provided that the propagation constant shift $\omega_a$ of waveguide W falls inside the tight-binding band of the array, as schematically shown in Fig.1(b). In this case, there are {\em two} values $k= \pm k_0$ of wave number $k$ that satisfy the resonance condition (16), corresponding to opposite group velocities $v_g$ in the lattice. Since $|G(k_0)|=|G(-k_0)|$, once the photon has tunneled from waveguide W into the array, it has the same probability to propagate either on the left or right sides of the array, i.e. excitation of light in the lattice is not chiral.  This result basically stems from the time-reversal symmetry of the Hamiltonian $\hat{H}$, which implies $\Omega(-k)=\Omega(k)$ and $G(-k)=G(k)$. Bidirectional excitation of the array in the $v=0$ regime is illustrated in Fig.3(a), along with the numerically-computed behavior of the survival probability $P_s(z)$ for the same parameter values of Fig.2. Note that a marked deviation of the decay from an exponential law is observed, with the appearance of damped Rabi-like oscillations. This result is mainly due to the strong coupling condition used in the numerical simulations (the ratio $\rho(0)/J$ is about 0.66), which is known  to induce damped Rabi-like flopping \cite{r27}. The bidirectional and symmetric excitation of the array is clearly observed from the symmetric discrete diffraction pattern at around $n=0$.\\
(ii) Let us now assume $v \neq 0$ and $v>0$ for the sake of definiteness. In particular, we focus our attention to the case where $v$ is larger than the critical value $v_c \equiv 2 J$, i.e. $\alpha>2 J d$. Under such a condition, it readily follows from Eq.(15) that $\Omega(k)$ is a monotonously decreasing function of $k$, and thus the resonance condition (16) is satisfied {\em solely} for one value $k=k_0$ of the wave number [Fig.1(c)]. This means that the photon in the array propagates this time {\it unidirectionally}, either on the right or left sides depending on whether $k_0<0$ or $k_0>0$, respectively. Figure 4 shows an example of chiral light excitation of the array for $k_0>0$. Such a chiral behavior has a simple physical interpretation, when we map the waveguide optics setup into the analogous problem of radiative decay of a two-level atom slowly drifting in a lattice of coupled cavities. In the reference frame of the moving atom, the array drifts at the speed $v$, so that time reversal symmetry of the bath Hamiltonian is broken and the dispersion relation of the bosonic e.m. modes of the array acquires in momentum space an additional ramp term $kv$ \cite{r26}, i.e. it is precisely described by Eq.(15). The resonance condition (16) merely corresponds to energy conservation of the atom-field system in the spontaneous emission process under the weak coupling approximation. Interestingly, as illustrated in Fig.1(c) the spontaneous emission process is not forbidden even though the frequency detuning $\omega_a$ falls far outside the tight-binding band $2 J$ of the lattice, namely the condition $|\omega_a|<2J$ for the spontaneous emission of the atom at rest ($v=0$) is now replaced by the condition
\begin{equation}
- \pi v - 2 J<\omega_a<\pi v - 2J.
\end{equation}
In the waveguide-optics problem of Fig.1(a), this means that light escape from waveguide W into the array can occur even though the propagation constant of the mode in waveguide W is far detuned from the one of the waveguides in the array, as if the bandwidth of the tight-binding array were effectively enlarged from $4J$ to $2 \pi v$.  An example of effective bandwidth enhancement due to waveguide tilt is illustrated in Fig.5 for parameter values which apply to femtosecond laser written waveguide arrays \cite{r2,r4,r25}.
%The figure depicts the numerically-computed survival probability $P_s$, after a propagation distance $z=10$ cm versus the waveguide detuning $\omega_a$ for parameter values as in Figs.3 and 4. 
In principle, according to Eq.(18) by increasing the tilting $v$ light can be coupled into the lattice for $|\omega_a|$ far beyond $4J$. However, since the effective coupling strength $\Gamma$ rapidly decreases as $v $  is increased [according to Eq.(17)] and the spectral coupling factor $|G_0|$ becomes highly asymmetric around $\omega_a=0$ [Fig.2(c)], in practice an effective bandwidth enhancement of 2-3 is obtained for typical parameter values, as shown in Fig.5. \par
\begin{figure}[htb]
\centerline{\includegraphics[width=8.7cm]{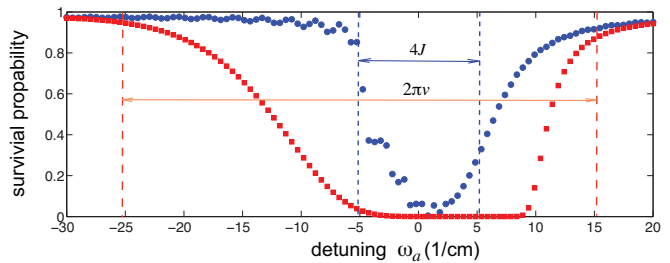}} \caption{ \small
(Color online) Numerically-computed survival probability $P_s$ in a waveguide lattice, after a propagation length $z=10$ cm,  versus waveguide detuning $\omega_a$ for the same parameter values as in Figs.3,4, Circles: $v=0$; squares: $\alpha= \pi /400$. The dashed vertical lines show the band of the static lattice ($v=0$) and of the moving lattice ($v= \alpha / 400$).}
\end{figure}

{\it Conclusions.} Chirality and effective band enlargement have been unravelled in waveguide-array evanescent mode coupling. Our results could be of relevance in applications, offering  a 
simple strategy for chiral excitation of light in a waveguide lattice and making light mode coupling more tolerant to  detuning effects arising, e.g., from fabrication imperfections.
On a more fundamental level, the  analogy between light escape dynamics and the spontaneous emission process of a two-level atom in a tight-binding continuum of bosonic modes  indicates that waveguide lattices, realized for example using the femtosecond laser technology \cite{r2,r4}, could provide an experimentally accessible platform to emulate the  exotic effect of chiral radiative emission of a moving quantum emitter without the use of gauge fields.\\

\par

{\bf The author declares no conflicts of interest.}

\end{document}